\documentclass[aps,prd,twocolumn,superscriptaddress,nofootinbib,amsmath,amssymb,floats,floatfix,nofrontpages,showkeys,longbibliography]{revtex4}

\usepackage[dvips]{graphicx,color}
\usepackage{subfig}
\usepackage{times}
\usepackage{mathrsfs}  
\usepackage{xcolor}
\usepackage[ colorlinks=true,
  urlcolor=blue,
  linkcolor=red,
  citecolor=blue
]{hyperref}

\RequirePackage{mathrsfs}
\RequirePackage{amsmath}
\RequirePackage{amssymb}
\RequirePackage{amsbsy}
\usepackage{graphicx}
\usepackage{multirow} 
\usepackage{caption}
\def\de#1/de#2{\frac{\partial {#1}}{\partial {#2}}}

\newcommand{\ba}{\begin{eqnarray}}
\newcommand{\ea}{\end{eqnarray}}
\newcommand{\be}{\begin{equation}}
\newcommand{\ee}{\end{equation}}

\usepackage{orcidlink}

\begin{document}

\title{Strange Quark Stars and Condensate Dark Stars in Bumblebee Gravity}

\author{Grigoris Panotopoulos \orcidlink{0000-0003-1449-9108}} \email[]{grigorios.panotopoulos@ufrontera.cl}
\affiliation{Departamento de Ciencias F{\'i}sicas, Universidad de la Frontera, Casilla 54-D, 4811186 Temuco, Chile.}


\author{Ali \"Ovg\"un \orcidlink{0000-0002-9889-342X}}
\email[Corresponding Author:]{ali.ovgun@emu.edu.tr}
\affiliation{Physics Department, Eastern Mediterranean University, Famagusta, Cyprus.}

\begin{abstract}
In this paper, we investigate the properties of relativistic stars made of isotropic matter within the framework of the minimal Standard Model Extension, where a Bumblebee field (BF) coupled to spacetime induces spontaneous Lorentz symmetry breaking. We adopt analytic equations-of-state describing either condensate dark stars or strange quark stars. We solve the structure equations numerically, and we compute the mass-to-radius relationships. The influence of the Bumblebee parameter $\mathbf{l}$ is examined in detail, and an upper bound is obtained using the massive pulsar (PSR) J0740+6620 and the strangely light High Energy Stereoscopic System (HESS) J1731-347 compact object.
\end{abstract}


\maketitle

\section{Introduction}

General Relativity (GR) and the Standard Model (SM) of Particle Physics, respectively, describe gravity classically and particles/interactions quantum mechanically. Unifying these theories remains a fundamental challenge. While proposed quantum gravity theories offer potential solutions, direct experimental verification is currently unattainable due to the requisite Planck scale ($10^{19}$ GeV) energies. Nonetheless, subtle quantum gravity effects might manifest at lower energy scales, opening avenues for experimental investigation. One such potential signal is the breaking of Lorentz symmetry, rooted in special relativity, ensures that physical laws are consistent for all observers in inertial frames. This symmetry, encompassing rotational and boost transformations, is fundamental to general relativity and the standard model of particle physics. In curved space-times, local Lorentz symmetry is maintained due to the Lorentzian nature of the background. However, violating Lorentz invariance introduces directional or velocity dependencies in physical variables, altering the dynamics of particles and waves. In 1989, Kostelecký and Samuel introduced Bumblebee gravity \cite{Kostelecky:1988zi}, a model for spontaneous Lorentz violation. In this model, a BF with a vacuum expectation value (VEV) breaks Lorentz symmetry through the dynamics of a single vector field, $B_\mu$. This approach suggests that investigating Lorentz symmetry violation could provide insights into Planck-scale physics, combining aspects of both General Relativity and the Standard Model within the framework of effective field theories known as the Standard Model Extension (SME), which offers a framework that includes all possible coefficients for Lorentz/CPT violation, with its gravitational sector defined on a Riemann-Cartan manifold, treating torsion and the metric as dynamic geometric quantities. While the SME allows for non-Riemannian terms, research has primarily focused on the metric approach, using the metric as the main dynamic field. There are certain effects of Lorentz violation within the gravitational sector have been explored in \cite{Kostelecky:2000mm,Kostelecky:2001mb,Bluhm:2004ep,Kostelecky:1989jw,AraujoFilho:2024ykw,Filho:2022yrk,Delhom:2022xfo,Delhom:2019wcm,Lambiase:2023zeo,Gullu:2020qzu,Li:2020dln,Maluf:2020kgf,Maluf:2014dpa,Casana:2017jkc,Bertolami:2005bh,Capozziello:2023tbo,Capozziello:2023rfv}.

\smallskip

New discoveries of super-dense objects are pushing the boundaries of our understanding in Astrophysics. While countless compact objects have been found, the exact laws governing their behavior remain a mystery. Recent observations have challenged the traditional theories, particularly general relativity and the standard model for neutron stars (NSs). One key example is the secondary object in the GW190814 merger \cite{Zhang:2020jmb,Zhang:2021iah,Gammon:2024gij}, \textcolor{black}{which has been proposed as a potential quark star. Its properties have been analyzed in the framework of general relativity, with models suggesting that they may be described by a single parameter ($\lambda$).} This theory aligns with recent research suggesting non-strange quark matter as the most stable form of baryonic matter under extreme conditions \cite{Holdom:2017gdc}. Promising findings have also been obtained by similar investigations using the quark matter phase or using other models for matter behavior (equation-of-state) \cite{Casana:2017jkc,Bertolami:2005bh,Capozziello:2023tbo,Capozziello:2023rfv,Gammon:2023uss}.

\smallskip

Strange quark stars are hypothetical compact objects proposed as an alternative to neutron stars \cite{Madsen:1998uh,Yue:2006it,Leahy:2007we,Arapoglu:2010rz,Eksi:2015red,Azri:2021pxu,Baym:2017whm,Pradhan:2024zzo,Bhar:2023yrf,Panotopoulos:2024iag,Rincon:2023ens,Rahaman:2011cw,Sedaghat:2022fue,Sedaghat:2024bnj,Deb:2017rhd,Bhar:2014tqa}. Composed of quark matter, which is theorized to be the absolutely stable ground state of hadrons \cite{Witten:1984rs,Farhi:1984qu}, these stars could explain the puzzling nature of super-luminous supernovae \cite{Ofek:2006vt,Ouyed:2009dr}. Unlike ordinary supernovae, these exceptionally bright explosions are 100 times more luminous and occur in approximately one out of every thousand cases.
\textcolor{black}{ According to the standard terminology, in the literature the names "quark stars", "strange stars" and "strange quark stars" are used to refer to the same object. Strange quark stars are made of u, d and s quarks, as the other three flavors are too heavy. Their mass is much larger than the chemical potentials involved, and so they are not relevant.} While quark stars remain theoretical at present, their existence cannot be definitively excluded. Some observed compact objects exhibit anomalous properties, such as unusually small radii, that standard neutron star models cannot fully explain \cite{Henderson:2007gu,Li:2011zzn,Maxim}. These discrepancies have fueled speculation about the potential presence of strange matter in these objects. While some researchers propose that strange matter could constitute the core of hybrid neutron stars \cite{Benic:2014jia,Yazdizadeh:2019ymr,EslamPanah:2018rfe}, others argue that such stars would be virtually indistinguishable from ordinary neutron stars \cite{Jaikumar:2005ne}.



\smallskip

The standard cosmological model posits that dark matter consists of weakly interacting massive particles (WIMPs). While this assumption aligns well with large-scale cosmic observations (scales of megaparsecs and larger) ($\geq Mpc$), it faces challenges at smaller galactic scales. A few discrepancies between observations and model predictions, such as the core-cusp problem, the diversity problem, the missing satellites problem, and the too-big-to-fail problem, arise \cite{Tulin:2017ara}. Self-interacting dark matter has been proposed as a potential solution \cite{Spergel:1999mh,Dave:2000ar}, as particle collisions within this model can mitigate the formation of sharp density cusps. Furthermore, if dark matter is composed of ultralight scalar particles with a mass below one electronvolt $m \leq eV$ and a weak repulsive self-interaction, it could form a Bose-Einstein condensate (BEC) \cite{harko1} with long-range correlations. This BEC scenario has been proposed as a potential explanation for the dark matter crisis on small (i.e. galactic) scales \cite{Tkachev:1986tr,Goodman:2000tg,Peebles:2000yy,harko2}.


\smallskip

Regarding the composition and inner structure of compact objects, the most massive pulsars \cite{Demorest, Antoniadis, recent} observed over the last 15 years or so are putting constraints on different equations-of-state, since any mass-to-radius relationship that predicts a highest mass lower than the observed ones must be ruled out. In order to model a certain star, it would be advantageous to know both its mass and its radius, which is not always the case as measuring the radius is way more difficult. There are some good strange quark star candidates, see e.g. Table 5 of \cite{Weber} or Table 1 of \cite{Maxim}, and also the recently discovered massive pulsar PSR J0740+6620 \cite{pulsar1, pulsar2, pulsar3} and the strangely light object HESS  \cite{hess}, where both the stellar mass and radius are known observationally.

\smallskip

Our current understanding of ultra-dense stars, particularly those exceeding the limitations of traditional neutron stars, remains a fascinating mystery in Astrophysics. \textcolor{black}{ Since compact objects are characterized by high matter densities and strong gravitational fields, they are excellent cosmic laboratories to constrain gravitational theories other than GR. Since the astrophysical implications of the Bumblebee gravity should also be investigated, in this work we propose to study 
the structural properties of compact stars made of isotropic matter, and in particular those of quark stars and self-gravitating spheres made of dark matter condensate \cite{harko3,harko4}.}

\smallskip

Our key objective is to explore the impact of a phenomenon known as spontaneous Lorentz symmetry breaking within these exotic stars. This breaking is induced by a hypothetical field, the BF, coupled to spacetime in the framework of the minimal SME.

 \smallskip

In the present article we propose to study non-rotating relativistic stars made of isotropic matter within bumblebee gravity adopting for the matter content two analytic equations-of-state. Our work is organized as follows: After this introductory section, we briefly review how to obtain interior solutions describing hydrostatic equilibrium within GR in the next section, while in section 3 we present the Bumblebee model as well as the corresponding modified structure equations. In the fourth section we comment on the equations-of-state assumed here, and we discuss our numerical results. Finally, we summarize and conclude our work in section 5. Throughout the manuscript we work in geometrical units setting the universal constants (Newton's constant and speed of light in vacuum) to unity, $G=1=c$, and we adopt the mostly positive metric signature $\{ -,+,+,+ \}$.

\section{Relativistic stars in General Relativity: Hydrostatic equilibrium and structure equations}

Due to their intense gravitational fields, the structure and dynamics of NSs are governed by Einstein’s equations of GR \cite{Einstein:1915ca}, given by:
\textcolor{black}{
\begin{equation}
    G_{\mu \nu} = \mathcal{R}_{\mu \nu} - \frac{1}{2} \mathcal{R} g_{\mu \nu} = 8\pi T_{\mu \nu}, 
\end{equation}}
where $R_{\mu \nu}$ and $R$ denote the Ricci tensor and Ricci scalar, respectively. The energy-momentum tensor for isotropic matter, $T_{\mu \nu}$, is expressed as:
\begin{equation}
    T_{\mu \nu} = P g_{\mu \nu} + (P+\rho) u_{\mu}u_{\nu},
\end{equation}
where $g_{\mu \nu}$ is the metric tensor, $P$ is the pressure, \textcolor{black}{$\rho$} is the energy density, and $u_{\mu}$  is the four-velocity. For static, spherically symmetric stars, the line element in Schwarzschild-like coordinates $\{ t,r,\theta,\phi \}$ is described as:
\textcolor{black}{
\begin{equation}\label{metric}
    ds^{2} = -e^{\nu(r)} dt^{2} + e^{\lambda(r)} dr^{2} + r^{2} (d\theta^{2} + \sin^{2} \theta d\phi ^{2}),
\end{equation}}
where $e^{\nu(r)}$ and $e^{\lambda(r)}$ are the metric functions. One obtains the Tolman-Oppenheimer-Volkoff (TOV) equations \cite{tolman,OV} for the equilibrium structure of NSs by solving the Einstein field equation with the above-defined metric,
\begin{equation}\label{tov1}
\frac{dP(r)}{dr}= -\frac{[\rho(r) +P(r)][m(r)+4\pi r^3 P(r)]}{r^2(1-2m(r)/r) } ,
\end{equation}
\begin{equation}\label{tov2}
\frac{dm(r)}{dr}= 4\pi r^2 \rho(r) .
\end{equation}
The metric functions become
\begin{equation}
    e^{\lambda(r)} = (1-2m/r)^{-1} ,
\end{equation}
\begin{align}
    \nu(r) &= \mathrm{ln(1-2M/R)} + 2\int_R^{r} dr' \frac{e^{\lambda(r')}}{r'^{2}} 
     [m(r')+4\pi r'^{3} P(r')] .
    \end{align}
Combined with the given the equation of state (EoS)---i.e., $P(\rho)$ of the matter---, TOV equations can be solved with the initial conditions at the center of the star, $m(r=0)$ = 0 and $\rho(r=0)$ = $\rho_c$, where $\rho_c$ is the central energy density. \textcolor{black}{ The stellar mass and radius are determined using the matching conditions at the surface of the star upon comparison to the Schwarzschild exterior solution in bumblebee gravity }\cite{Schwarzschild:1916uq}
\begin{equation}
ds^2 = -(1-2M/r) \: dt^{2}+(1+\mathbf{l})\frac{1}{1-2 M/r} \: dr^{2}+r^{2}d\Omega^{2}.
\end{equation}
Thus, the radius of the star is determined by requiring that the energy density vanishes at the surface, $P(R)$ = 0, and the stellar mass is then given by $M = m(R)$.

\section{Bumblebee gravity: Modified TOV equations
	\label{sec2}}

In this section, we briefly review the bumblebee gravity, a theory that expands on GR. Inspired by the works of Kostelecky and collaborators \cite{Kostelecky:1988zi}, bumblebee gravity introduces a twist: it breaks a fundamental symmetry (Lorentz symmetry) within the realm of gravity. This twist manifests as a special value (nontrivial vacuum expectation value) that influences how other fields behave around a mysterious "bumblebee field". Interestingly, even with these extra interactions, bumblebee gravity preserves the geometrical framework and basic laws established by GR in curved spaces. 


Among various models capable of breaking Lorentz symmetry, one of the simplest approaches involves a vector field \( B^\mu \), known as the BF, within a torsion-free spacetime. This can be expressed as follows \cite{Casana:2017jkc}:
\begin{widetext}
\begin{equation}\label{BumblebeeAction}
	\begin{split}
		\mathcal{S} =& \int \mathrm{d}^{4}x\sqrt{\left|g\right|}
		\left(
		\frac{1}{16\pi}\left(\mathcal{R} + \xi B^{\mu}B^{\nu}\mathcal{R}_{\mu\nu}\right)-\frac{1}{4}B_{\mu\nu}B^{\mu\nu}-V\left(B^{\mu}B_{\mu}\pm b^{2}\right)+\mathcal{L}_{M}[g]
		\right),
	\end{split}
\end{equation}
 which encapsulates the interaction between the bumblebee vector field (\(B_{\mu}\)) and gravity, represented by the Ricci tensor (\(R_{\mu\nu}\)). The strength of this interaction is modulated by a coupling constant (\(\xi\)). Notice that in this work we assume that the Lagrangian of matter content only depends on the metric tensor $g_{\mu \nu}$, and it is $B_\mu$ independent.
\end{widetext}


\textcolor{black}{The BF $B_{\mu\nu}\equiv \partial_{\mu}B_{\nu}-\partial_{\nu}B_{\mu}$ does not coupled  with matter, as outlined by the matter Lagrangian density (\(L_m\)) which contains a perfect fluid.} The potential function \(V(B_{\mu}B_{\nu} \pm b^2)\), where \(b^2\) is a positive constant, dictates the behavior of the BF. Notably, this potential enables the field to achieve a non-zero vacuum expectation value, specifically \(B_{\mu} = b_{\mu} = (0, b_{r}(r), 0, 0)\), where \(b_{r}(r)\) is a function of the radial coordinate.


To break a specific symmetry, such as $U(1)$, the potential must reach a minimum at  $V=0$, with its derivative $V'=0$ also being zero at that point. This condition allows the BF to acquire a constant, non-zero vacuum expectation value, denoted by $b_{\mu}$, which satisfies ${B^{\mu}} \equiv b^{\mu}$ with $b^{\mu}b_{\mu} = \mp b^{2}\equiv $ constant (note that $b^{2}$ is a real positive constant).


Varying the action, as given in Eq. \eqref{BumblebeeAction}, we derive the equations governing both the gravitational field and the dynamics of the BF, including the corresponding equations of motion.
\begin{equation}
\label{GravitationalﬁeldEq}
	\mathcal{R}_{\mu\nu} -\frac{1}{2} g_{\mu\nu} \mathcal{R} = 8 \pi \,\left(T_{\mu\nu}^{B}+T_{\mu\nu}^{M}\right),
\end{equation}
\begin{equation} \label{eqn11}
\nabla_\mu B^{\mu\nu} = \mathcal{J}_B^\nu + \mathcal{J}_M^\nu.
\end{equation}

\textcolor{black}{In the Bumblebee gravity framework, two essential terms are: the matter source term associated with the Bumblebee field, represented by \(\mathcal{J}^{M}_{\nu}\), and the self-interaction current of the Bumblebee field (BF), denoted by \(\mathcal{J}^{B}_{\nu} = 2 V' B^{\nu} - (\varrho /8 \pi) B_{\mu}\mathcal{R}^{\mu\nu}\).}

\textcolor{black}{Given that the field strength tensor vanishes, \(B_{\mu\nu}=0\), Eq. \ref{eqn11} yields the relation \(\mathcal{J}^{\mu}_{B} = - \mathcal{J}^{\mu}_{M}\). We assume that \(|\mathcal{J}^{\mu}_{B}|\approx 0\), indicating negligible interaction. Therefore, it is reasonable to assume the absence of coupling between the matter sector and the Bumblebee field. This assumption is further supported by the fact that the matter Lagrangian \(\mathcal{L}_m\) in our model describes a perfect fluid. This scenario aligns with the approach previously considered in Ref. \cite{Neves:2024ggn}.}

The prime symbol (\('\)) denotes differentiation with respect to the potential argument, specifically \(V' \equiv \partial V(y)/\partial y\), where \(y = B^{\mu} B_{\mu} \pm b^{2}\).



\begin{widetext}
 
Additionally, $T^{B}_{\mu\nu}$  represents the energy-momentum tensor associated with the BF, which can be expressed as

\begin{equation}\label{EnergyMomTens}
	\begin{split}
		T^{B}_{\mu\nu}& =B_{\mu\alpha}B^{\alpha}_{\,\,\,\nu}-\frac{1}{4}g_{\mu\nu}B^{\alpha\beta}B_{\alpha\beta}-g_{\mu\nu}V+2B_{\mu}B_{\nu}V'+\frac{\varrho}{8\pi}
		\left[\frac{1}{2}g_{\mu\nu}B^{\alpha}B^{\beta}\mathcal{R}_{\alpha\beta}\right.\\
		&\left.-B_{\mu}B^{\alpha}\mathcal{R}_{\alpha\nu}-B_{\nu}B^{\alpha}\mathcal{R}_{\alpha\mu}+\frac{1}{2}\nabla_{\alpha}\nabla_{\mu}\left(B^{\alpha}B_{\nu}\right)+\frac{1}{2}\nabla_{\alpha}\nabla_{\nu}\left(B^{\alpha}B_{\mu}\right)\right.\\
		&\left.-\frac{1}{2}\nabla^{2}\left(B_{\mu}B_{\nu}\right)-\frac{1}{2}g_{\mu\nu}\nabla_{\alpha}\nabla_{\beta}\left(B^{\alpha}B^{\beta}\right)
		\right].
	\end{split}
\end{equation}
\end{widetext}

Recall that $T_{\mu\nu}^{M}$ represents the energy-momentum distribution of matter, \textcolor{black}{while $\varrho$ is the real coupling constant (with mass dimension $-1$) that governs the non-minimal interaction between gravity and the bumblebee field.} Additionally, we introduce a constant value denoted by $b^{\mu}b_{\mu} = \mp b^{2}\equiv \text{constant}$ related to the BF (where $b_{\mu}$ is the VEV and b is its magnitude). These factors influence the radial component $b_r(r)$ of the BF when it reaches its minimum energy state (VEV). Interestingly, research by Casana et al. \cite{Casana:2017jkc} shows that this radial component can be expressed as  $b_{r}(r) = |b| e^{\zeta(r)}$.


The next step is to understand how this VEV of the BF affects the geometry of spacetime itself. This is described by the metric, which will be expressed in the following spacetime
\begin{equation}\label{Bumbelbeemetric2}
	ds^2 = -e^{2\chi (r)}dt^{2}+e^{2\zeta(r)}dr^{2}+r^{2}d\Omega^{2},
\end{equation}
where $\chi(r)$ and $\zeta(r)$, valid in the region $0 \leq r \leq R$, represent the sought metric functions. It is noteworthy that choosing $e^{-2\zeta(r)} = \text{g}(r)$ is always possible, resulting in:
\begin{equation}\label{MetCoeffFunc-g1}
	\begin{split}
		g (r) =1-\frac{2\, M(r)}{r},
	\end{split} 
\end{equation}
in which the collective mass function denoted by $\mathrm{M}(r)$ can be viewed as the combination of the black hole (BH) mass.


Thus, for an interacting system of this nature, the energy-momentum tensor associated with the spacetime metric can be represented as a perfect fluid $T^{\mu}_{\,\,\nu} = \text{diag}\left[-\rho(r),P_{r}(r),P_{\theta}(r),P_{\phi}(r)\right]$. \textcolor{black}{Consequently, these considerations facilitate the derivation of the Einstein field equations $G_{\mu\nu} = 8\pi T_{\mu\nu}$ as }

\begin{widetext}

\begin{subequations}\label{EinsteinFieldEqs1}
	\begin{align}
   8\pi \rho(r) & =  \frac{e^{-2 \zeta(r)}}{r^{2}}\left[e^{2 \zeta(r)}-(1+\mathbf{l})\left(1-2 r \zeta(r)^{\prime}\right)\right]=\kappa \rho  \\
8\pi P_{r}(r)& =\frac{e^{-2 \zeta(r)}}{r^{2}}\left[(1+\mathbf{l})\left(1+2 r \chi(r)^{\prime}\right)-e^{2 \zeta(r)}-\ell r^{2}\left(\chi(r)^{\prime \prime}+\chi(r)^{\prime 2}-\chi(r)^{\prime} \zeta(r)^{\prime}-\frac{2}{r} \zeta(r)^{\prime}\right)\right]
\\ 8\pi P_{\theta}(r)= 8\pi P_{\varphi}(r)&=(1+\mathbf{l}) e^{-2 \zeta(r)}\left[\chi(r)^{\prime \prime}+\chi(r)^{\prime 2}-\chi(r)^{\prime} \zeta(r)^{\prime}+\frac{1}{r}\left(\chi(r)^{\prime}-\zeta(r)^{\prime}\right)\right].
        \label{EinsteinFieldEqs1-34}
	\end{align}
\end{subequations}
\end{widetext}

\textcolor{black}{Within the bumblebee gravity framework, parameter of Lorentz symmetry breaking is $\mathbf{l}$, which itself is linked to the coupling strength $\xi$ of the BF and a constant related to the BF $b²$ as with $\mathbf{l}=\xi b^2$.} The prime symbol $(')$ denotes differentiation with respect to the radial position (r). Interestingly, when we take the limit as $\mathbf{l}$ approaches zero,  has a significant implication.  In this specific limit ($\mathbf{l} \rightarrow 0$ ), the Einstein field equations (refer to Eq. \eqref{EinsteinFieldEqs1} for reference) simplify back to their standard form. 


Now, let us bring in another piece of the puzzle: the equation of state $P_{r}(r) = \omega \rho(r)$, which relates pseudo-pressure ($P_{r}(r)$) and energy density ($\rho(r)$) with a constant factor ($\omega$). By combining this equation of state with the modified Einstein field equations we obtained earlier (Eq. \eqref{EinsteinFieldEqs1-34}), and considering a well-known law in physics (conservation of energy-momentum) $T_{\mu;\nu}^{\nu} = 0$, we can arrive at a modified version of the TOV equations.\textcolor{black}{These modified TOV equations are crucial for understanding how bumblebee gravity affects the structure and stability of stars and other self-gravitating objects} \cite{Neves:2024ggn}:


\begin{subequations}
\begin{align}
\frac{dM(r)}{dr}
  &= 4\pi r^2\,\rho(r),
  \label{TOV2}\\
\begin{split}
\frac{dP_{r}(r)}{dr}
  &= -\Bigl(\frac{\rho(r)+P_{r}(r)
    +\ell[\rho(r)+2P_{r}(r)]}
    {1+2\ell}\Bigr)\,\chi'(r)\\
  &\quad
    -\,\frac{\ell\bigl[8\pi r\,P_{r}(r)
    - M''(r)\bigr]}
    {(1+2\ell)\,8\pi r^{2}},
\end{split}
  \label{TOV3}
\end{align}
\end{subequations}

and
\begin{equation}
\chi'(r)=\frac{(1+2 \mathbf{l}) 8 \pi r^{3} P_{r}(r)+(2+3 \mathbf{l}) M(r) -\mathbf{l} r M'(r)}{(2+3 \mathbf{l}) r(r-2 M(r))}.
\end{equation}
Hence, in the exterior (vacuum) region where \( T_{\mu \nu} = 0 \), the geometry simplifies to the bumblebee BH solution \cite{{Casana:2017jkc}}
\begin{equation}
ds^2 = -(1-2M/r) \: dt^{2}+\frac{\mathbf{l}+1}{1-2 M/r} \: dr^{2}+r^{2}d\Omega^{2},
\end{equation}
{\color{black} or, after a coordinate transformation and a stellar mass redefinition as follows
\begin{equation}
t \rightarrow t, \; r \rightarrow \sqrt{1+l} \: r, \; M \rightarrow \sqrt{1+l} \: M,
\end{equation}
it may be recast in the following form \cite{Capozziello:2023tbo}
\begin{equation}ds^2 = -(1-2M/r) \: dt^{2} + \frac{1}{1-2 M/r} \: dr^{2} + q^2
\: r^{2} d\Omega^{2},
\end{equation}
where $q^2=1/(\mathbf{l}+1)$, and where now the modification due to the Bumblebee gravity is relocated to the angular part of the line element.
In the discussion to follow, however, we shall employ the more standard coordinate system, namely the previous form of the metric tensor without the coordinate transformation and the mass redefinition.}

\textcolor{black}{This result indicating that a background field carrying Lorentz-violating effects also slightly deforms spacetime. Despite these effects being significantly small, with a sensitivity at the $10^{-13}$ level, spacetime is described as approximately flat. This smallness has facilitated the computation of upper bounds on a parameter $\mathbf{l}$, referenced in \cite{Casana:2017jkc}.}

When studying a star in Bumblebee gravity, we need to ensure our solutions for the different regions (interior and exterior) seamlessly match at the star's surface, located at a specific radius ($R$). To achieve this perfect match, we impose three matching conditions as follows:
\begin{itemize}

\item \textbf{Matching Mass:} 

\begin{equation}  
M=M(R)
\end{equation} 
the first condition states that the value of the mass function ($M(r)$) at the surface ($r = R$) must be equal to the total mass ($M$) of the star.  

\item \textbf{Zero Pressure at the Surface:} 
    
\begin{equation}  
P_r(R)=0,
\end{equation} 
the second condition requires the pressure at the surface to be zero. This condition allows us to compute the radius of the star.

\item \textbf{Setting the Spacetime Geometry:} 

\begin{equation}
e^{2 \chi(R)} = 1-2M/R,
\end{equation} 
the third condition involves the exponential term in a metric function ($e^{2 \chi(R)}$). Here, we stipulate that its value at the surface ($r = R$) is related to the star's mass ($M$) and radius ($R$) through a specific relationship ($1 - 2M/R$). This essentially sets the initial condition for this metric function, which helps describe the geometry of spacetime around the star.
\end{itemize}

In summary, the aforementioned matching conditions act as bridge equations, ensuring that our solutions for the star's internal structure and the surrounding vacuum smoothly connect at the boundary, providing a physically consistent description of a Bumblebee gravity star.


\begin{figure}[h!]
\centering
\includegraphics[width=0.5\textwidth]{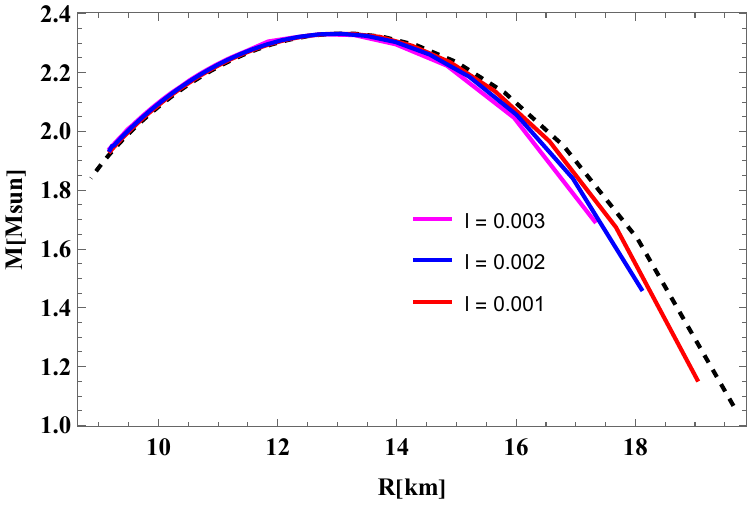} \\
\includegraphics[width=0.5\textwidth]{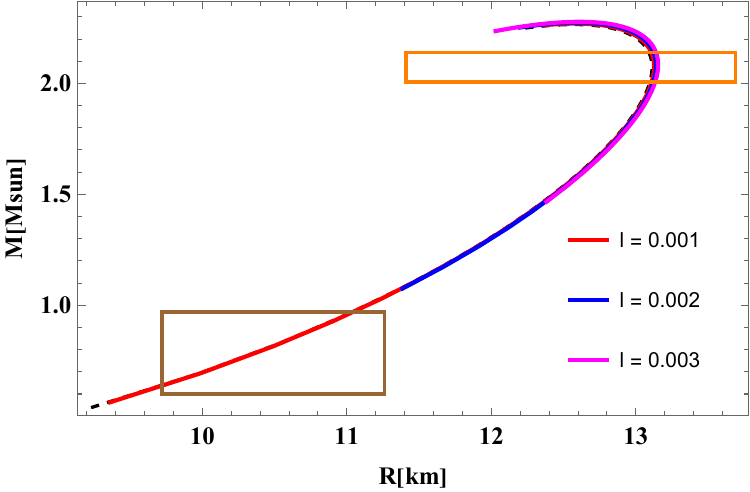}
\caption{
Mass-to-radius relationships for the two EoSs considered here. Upper panel corresponds to polytrope (l$=0.001,0.002,0.003$ from top to bottom), lower panel to the extreme MIT bag model (l$=0.001,0.002,0.003$  from left to right). The contours in orange and brown color indicate the allowed mass and radius range of the massive pulsar PSR J0740+6620 and the light HESS J1731-347 compact object, respectively.
}
\label{fig:1} 	
\end{figure}

\begin{figure}[h!]
\centering
\includegraphics[width=0.5\textwidth]{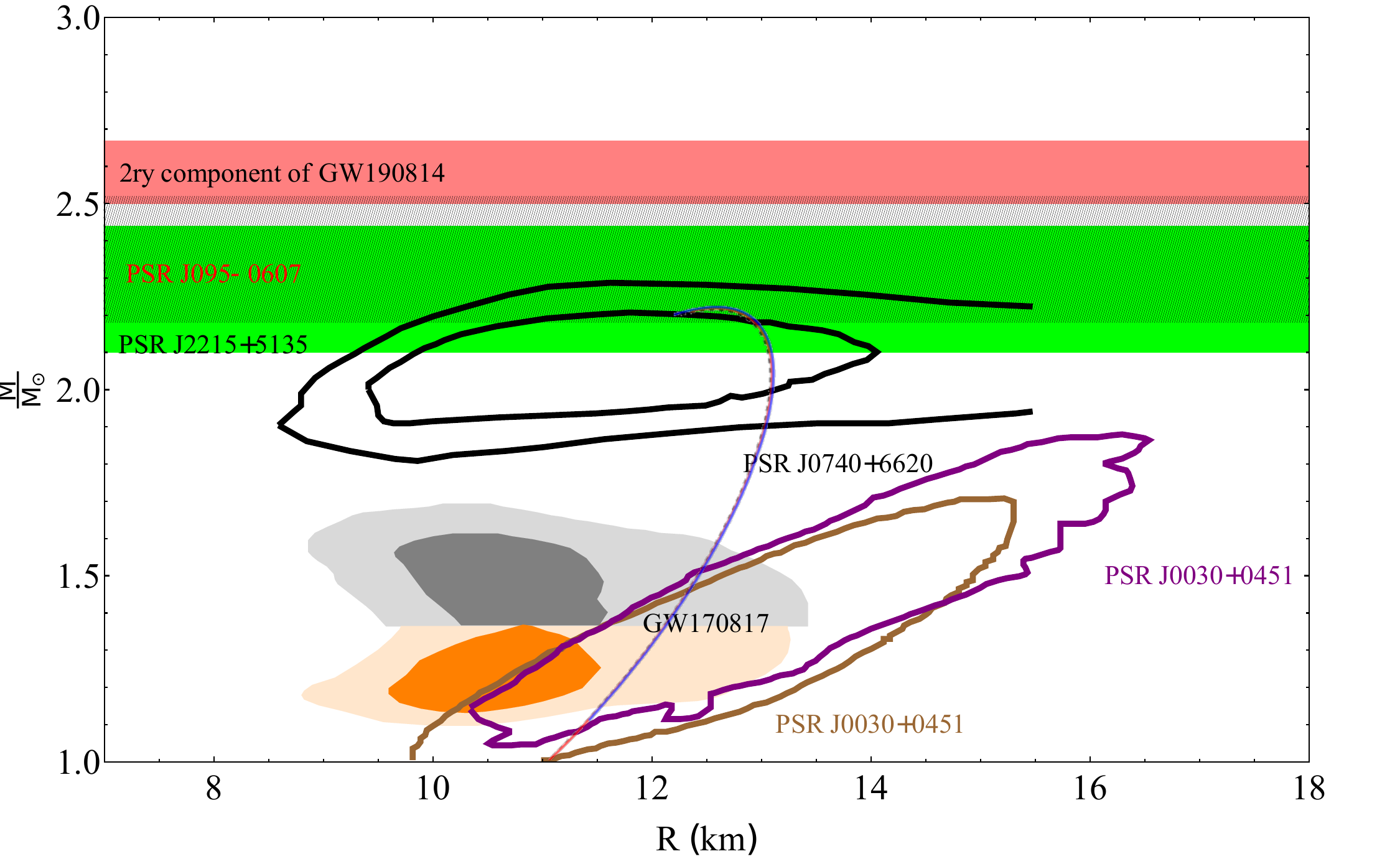}
\caption{ Mass-to-radius relationships for the MIT bag model considered here. Figure corresponds to the extreme MIT bag model (l$=0.001,0.002,0.003$ from left to right). 
The gray and orange regions depict the mass–radius constraints derived from the GW170817 event. The black region corresponds to pulsar J0740+6620, while the green and black hatched areas represent pulsars J2215+5135 \cite{Linares:2018ppq} and PSR J095-0607, respectively. The red region indicates the secondary component of GW190814. Additionally, the brown and purple regions illustrate two distinct reports on the mass–radius measurements of pulsar J0030+0451 \cite{Miller:2019cac,Riley:2019yda}.
}
\label{fig:2a} 	
\end{figure}


\begin{figure}[h!]
\centering
\includegraphics[width=0.5\textwidth]{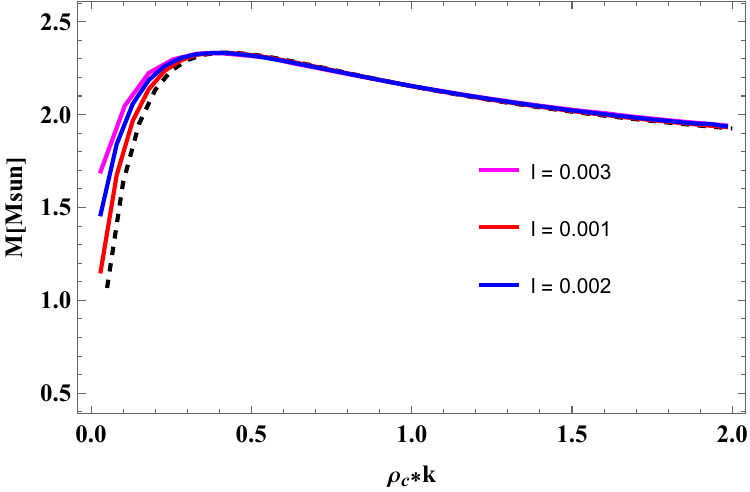} \\
\includegraphics[width=0.5\textwidth]{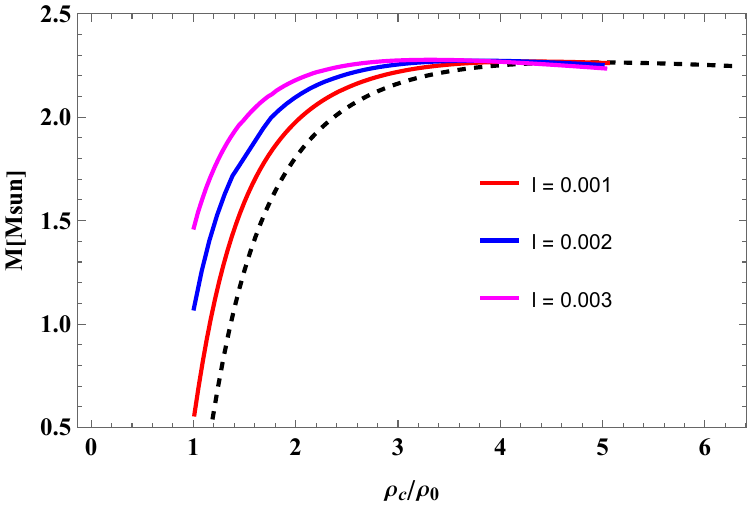}
\caption{
Stellar mass versus normalized central energy density ($\rho_c/\rho_0$ in the case of the MIT bag model and $\rho_c \: k$ in the case of the polytrope) for the two EoSs considered here. Upper panel corresponds to polytrope, lower panel to the extreme MIT bag model. 
}
\label{fig:2} 	
\end{figure}


\begin{figure}[h!]
\centering
\includegraphics[width=0.5\textwidth]{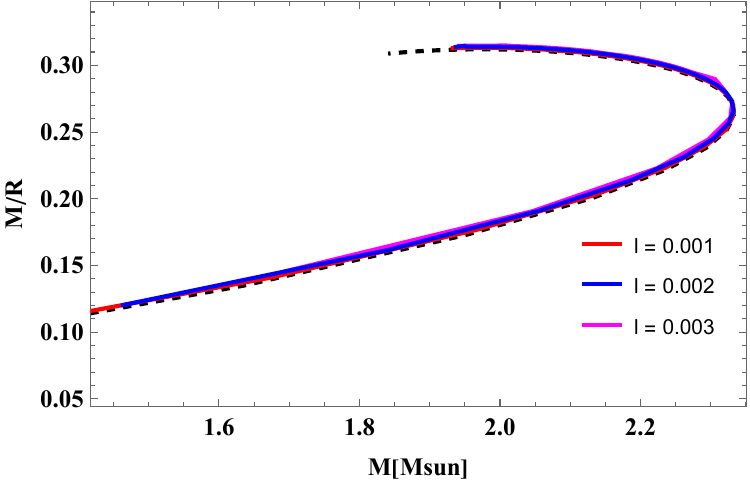} \\
\includegraphics[width=0.5\textwidth]{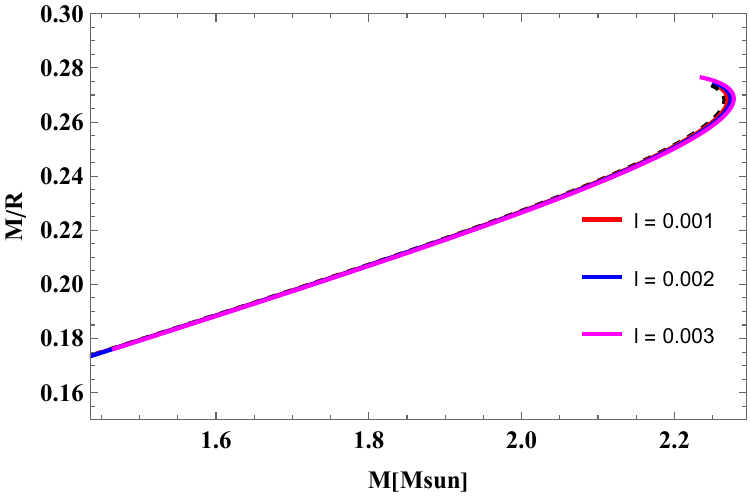}
\caption{
Factor of compactness versus stellar mass for the two EoSs considered here. Upper panel corresponds to polytrope, lower panel to the extreme MIT bag model.
}
\label{fig:3} 	
\end{figure}


\section{Properties of stars: Numerical analysis and discussion of main results}
\label{Sec3}

In this section, we explore the properties of relativistic stars made of isotropic matter, $P_r=P_\theta=P_\phi=P$, within the framework of Lorentz-violating gravity theories. Through numerical integration of the relevant structure equations, we subsequently present and analyze our key findings.


\subsection{Equation-of-state}

In the discussion to follow we shall consider i) a polytropic EoS and ii) Massachusetts Institute of Technology (MIT) bag model \cite{Chodos:1974je,Chodos:1974pn} for quark matter. 

On the one hand, polytropes of the form
\begin{equation}
P = k \rho^\gamma, \; \; \; \; \; \gamma=1+\frac{1}{n}
\end{equation}
\textcolor{black}{with $n$ being the polytropic index, are used to describe interior solutions of white dwarfs (index $n=3$ in the ultra-relativistic limit or $n=3/2$ in the non-relativistic limit) as well as condensate dark stars (index $n=1$) \cite{harko3,harko4}.  In the latter case, the constant $k$ depends on the mass of the dark matter particle, $m_B$, and on the scattering length, $L$, as follows \cite{harko1,harko4}}
\begin{equation}
k = \frac{2 \pi L}{m_B^3}
\end{equation}
In the following we shall consider the case $n=1$ assuming \textcolor{black}{$k=4.012 \times 10^{-4}\, \mathrm{fm}^3 / \mathrm{MeV}$}. 

\textcolor{black}{On the other hand, within the MIT bag model there is one class of models that is characterized by analytic, linear EoSs of the form
\begin{equation}
P = a \: (\rho - \rho_0) 
\end{equation}
where the values of the surface energy density, $\rho_0$, and of the dimensionless numerical coefficient, $a$, depend on the values of the three parameters of the MIT bag model, namely the bag constant, $B$, the mass of the s quark, $m_s$, and the strong coupling constant, $\alpha_c$. In particular, in the case of the extreme Strange Quark Star(SQSB40) MIT bag model \cite{Gondek-Rosinska:2008zmv} the numerical values of the three parameters of the MIT bag model are as follows: The bag constant \textcolor{black}{$B=40  \, \mathrm{MeV}/\mathrm{fm}^3$, while the mass of the s quark and the strong coupling constant are set to $m_s=100 \, \mathrm{MeV}$ and $\alpha_c=0.6$ \cite{Gondek-Rosinska:2008zmv} respectively,} and so $a=0.324$ and \textcolor{black}{$\rho_0=3.0563 \times 10^{14}  \, \mathrm{g}/\mathrm{cm}^3$ } \cite{Gondek-Rosinska:2008zmv}. }


\textcolor{black}{Neves and Gardim \cite{Neves:2024ggn} investigated quark stars in bumblebee gravity using the simplest version of the MIT bag model, characterized by massless strange quarks $m_s = 0$ and absence of strong interactions $\alpha_c = 0$. In contrast, our analysis utilizes a more sophisticated MIT bag model (extreme MIT bag model), incorporating finite strange quark mass ($m_s = 100$ MeV) and a significant strong coupling constant ($\alpha_c = 0.6$). These additional physical ingredients substantially influence the resulting mass-radius (M-R) relations. Explicitly, our generalized EoS leads to a systematically more compact stellar structure, as evident from a pronounced shift of the M-R curves towards smaller radii at comparable masses. Furthermore, the maximum mass predicted by our model also deviates significantly from that of Neves and Gardim \cite{Neves:2024ggn}, underscoring the importance of quark mass and interaction terms in accurately modeling quark star properties. These observations indicate that even subtle variations in the assumed quark EoS yield clear and measurable differences in the structural predictions of quark stars within the bumblebee gravity framework.}

\textcolor{black}{On the other hand, Ji et al. \cite{Ji:2024aeg} performed a comprehensive numerical analysis of neutron stars in the context of bumblebee gravity using various neutron star EoSs. They concluded that altering the neutron star EoSs does not result in qualitative changes to the M-R relations. While this assertion holds true for neutron star matter, which generally exhibits similar physical characteristics across different EoS models, it does not necessarily extend to quark matter. Our analysis explicitly reveals a contrasting scenario: the qualitative and quantitative stellar properties of quark stars are highly sensitive to the detailed EoS assumptions, due to fundamental differences in the state of matter (quark matter versus neutron matter). Consequently, the M-R curves in our analysis, characterized by the extreme MIT bag model, are qualitatively distinct from those obtained by Neves and Gardim (simplified MIT bag model) \cite{Neves:2024ggn} and Ji et al. (neutron star EoSs) \cite{Ji:2024aeg}. }

\textcolor{black}{In our analysis, we have systematically explored the effects of varying the bumblebee parameter ($\mathbf{l}$), which quantifies the strength of Lorentz symmetry breaking. While changing this parameter indeed modifies stellar masses and radii quantitatively, our findings indicate that the qualitative differences between the M-R curves primarily arise from the selected EoS rather than the specific value of $\mathbf{l}$. Thus, the primary factor distinguishing our results from previous analyses in the literature \cite{Neves:2024ggn,Ji:2024aeg,chinos} is the choice and sophistication of the EoS rather than variations in the Lorentz-violating parameter.}

\textcolor{black}{Hence, our comparative analysis highlights that the qualitative differences in the mass-radius relations observed across various studies in the bumblebee gravity context predominantly stem from the assumptions regarding the EoS.}

\subsection{Numerical results}

In Fig. \ref{fig:1}, we show the mass-to-radius relationships for the two EoSs considered here, i.e. a polytrope with $n=1$ (upper panel) and the extreme MIT bag model (lower panel). The dashed curves represent the standard mass-to-radius relationship within GR. Those curves serve as a reference for how quark stars behave under GR without modifications from the bumblebee gravity theory. {\color{black} Since in previous works it has been shown that the Bumblebee parameter must be very small, we have considered here the following three different numerical values of the $\mathbf{l}$ parameter, namely $\mathbf{l}=0.001$ in red color, $\mathbf{l}=0.002$ in blue, and $\mathbf{l}=0.003$ in magenta.}

{\color{black} In both cases (polytrope and MIT bag model),} the mass-to-radius relationships show slight deviations from the GR case. The introduction of a small bumblebee parameter indicates minor changes in the structure and stability of the stars considered here, reflecting how {\color{black} compact relativistic stars} might behave under weak modifications to gravity. Assuming a larger bumblebee parameter, the mass-to-radius relationship exhibits {\color{black} slightly more significant deviations than the case of GR.} This indicates that stronger modifications to gravity {\color{black} might} have a notable impact on the structure, stability, and maximum mass of quark stars.

{\color{black}Figure \ref{fig:2a} also presents the mass–radius constraints inferred from the GW170817 event. The black area corresponds to pulsar J0740+6620, while the green and black hatched regions depict pulsars J2215+5135 \cite{Linares:2018ppq} and PSR J0952-0607, respectively. The red area highlights the secondary component of GW190814. Furthermore, the brown and purple regions represent two separate studies providing mass–radius measurements for pulsar J0030+0451 \cite{Miller:2019cac,Riley:2019yda}.}

\smallskip

In Fig. \ref{fig:2} the stellar mass versus central (normalized and dimensionless) energy density is shown. As in previous figure, the upper panel corresponds to the polytrope, while the lower panel to the extreme MIT bag model. The dashed curves correspond to GR, the blue curves to the lowest value of $\mathbf{l}$, and the red curves to the highest value of $\mathbf{l}$. The curves increase, reach a maximum value and then they decrease. The maximum stellar mass is the one that is shown in Fig. \ref{fig:1}. The Harrison-Zeldovich-Novikov criterion \cite{Harrison,Zeldovich} for stability
\begin{equation}
\frac{d M}{d \rho_c} > 0
\end{equation}
states that a stellar model is a stable configuration only if the mass of the star grows with the central energy density. Therefore according to the criterion, only the first part of the function $M(\rho_c)$ is physical, namely up to the maximum stellar mass.

\smallskip

Next, in Fig. \ref{fig:3} we display the factor of compactness, $M/R$, as a function of the stellar mass for the two EoS assumed in this work. Panels and colors are the same as before. The factor of compactness increases with the stellar mass, which acquires a maximum value, and which is precisely the same that is shown in Figures  \ref{fig:1} and \ref{fig:2}. Notice that the factor of compactness satisfies the Buchdahl limit of GR, $(M/R) \leq 4/9 \approx 0.444$ \cite{Buchdahl:1959zz}.

\section{Conclusion}\label{sec7} 

To summarize our work, in the present article we investigated the properties of non-rotating relativistic stars made of isotropic matter within the bumblebee gravity in four-dimensional space-time. For the matter content of the stars we considered analytic equations-of-state corresponding to quark matter (extreme MIT bag model) and condensate dark stars (polytropic EoS with index $n=1$). In the first part of the article we presented the exterior vacuum solution, the structure equations describing hydrostatic equilibrium of interior solutions as well as the appropriate conditions both at the center and at the surface (matching conditions) of the stars.

\smallskip

Next, we integrated numerically the generalized TOV equations imposing the initial conditions at the center of the star, and then making use of the matching conditions we computed the stellar mass and radius as well as the factor of compactness. The impact of the bumblebee parameter $\mathbf{l}$ on the properties of the stars was shown in three figures, where we displayed i) the stellar mass versus stellar radius, ii) the factor of compactness as a function of the stellar mass, and iii) stellar mass versus central energy density. 

\smallskip

{\color{black}Our results show that a) as we increase $\mathbf{l}$ the deviation from GR becomes more significant, as expected, b) the factor of compactness satisfies the Buchdahl limit of GR, $(M/R) \leq 4/9 \approx 0.444$ \cite{Buchdahl:1959zz}, c) regarding the MIT bag model, increasing $\mathbf{l}$ shifts the M-R relationships to the right, which implies that at a certain point the EoS will not be able to model the HESS compact object, d) regarding the polytrope, increasing $\mathbf{l}$ shifts the M-R profiles downwards, which implies that at a certain point the EoS will not be capable of supporting massive stars at two solar masses. 
In principle, raising the numerical value of the Bumblebee parameter would allow one to establish an upper bound on $\mathbf{l}$. However, since this constraint would be weaker than existing limits in the literature, we have not pursued it in our analysis.}

\section{ACKNOWLEDGMENTS}

\textcolor{black}{ We wish to thank the anonymous reviewers for useful comments and suggestions.} A.{\"O}. would like to acknowledge the contribution of the COST Action CA21106 - COSMIC WISPers in the Dark Universe: Theory, astrophysics and experiments (CosmicWISPers), the COST Action CA21136 - Addressing observational tensions in cosmology with systematics and fundamental physics (CosmoVerse), the COST Action CA22113 - Fundamental challenges in theoretical physics (THEORY-CHALLENGES), the COST Action CA23130 - Bridging high and low energies in search of quantum gravity (BridgeQG) and the COST Action CA23115 - Relativistic Quantum Information (RQI) funded by COST (European Cooperation in Science and
Technology). We also thank EMU, TUBITAK, ULAKBIM, Turkiye and SCOAP3, Switzerland for their support.



\end{document}